\documentclass[11pt,a4paper]{article}
\usepackage[hyperref]{rocling2026}
\usepackage[T1]{fontenc}
\usepackage{times}
\usepackage{latexsym}

\usepackage{microtype}
\usepackage{amsmath,amsfonts,amssymb}
\usepackage{graphicx}
\usepackage{booktabs}
\usepackage{array}
\usepackage{enumitem}
\usepackage{stfloats}
\setlist[itemize]{leftmargin=*, labelindent=0pt, itemsep=0pt, topsep=0pt, parsep=0pt, partopsep=0pt}
\newcolumntype{L}[1]{>{\raggedright\arraybackslash}m{#1}}
\newcolumntype{C}[1]{>{\centering\arraybackslash}m{#1}}
\roclingfinalcopy 

\title{GTCRN-SS: Lightweight Causal Speech Separation with\\Residual Allocation and Reconstruction Refinement}

\author{
  \textbf{Shao-Chun Hu\textsuperscript{1,*}},
  \textbf{Zi-Xiang Lin\textsuperscript{1,*}},  
  \\
  \textbf{Hsin-Min Wang\textsuperscript{2}},
  \textbf{Jeih-Weih Hung\textsuperscript{1}},
  \textbf{Hung-Shin Lee\textsuperscript{3}}
  \\
  \\
  \textsuperscript{1}Dept. of Electrical Engineering, National Chi Nan University
  \\
  \textsuperscript{2}Inst. of Information Science, Academia Sinica
  \\
  \textsuperscript{3}Grad. Inst. of AI Interdisciplinary Applied Technology, National Taiwan Normal University
  \\
  \textsuperscript{*}Equal contribution
}

\date{}

\begin{document}
\maketitle

\begin{abstract}Low-complexity causal speech separation must preserve speaker information and allocate residual mixture energy without relying on future feature frames. We present GTCRN-SS, a two-stage time-frequency separator that adapts the compact GTCRN enhancement backbone to fixed-output speech separation. The model combines internal residual allocation, reconstruction-conditioned refinement, and frame-aligned recurrent-state transfer. Its complex-mask estimator assigns non-target content to an unreturned residual branch, while the complementary speech estimate conditions a second stage whose final masks remain anchored to the original mixture. A lightweight low-rank adapter transfers causal recurrent evidence between recursive executions without adding another recurrent memory network. The neural graph excludes future feature frames and incurs only the fixed analysis look-ahead of the centered short-time Fourier transform. On noisy Libri2Mix, GTCRN-SS achieves 10.1~dB SDRi and 9.5~dB SI-SNRi with fewer than one million parameters and a low computational footprint. Independently trained ablations confirm the benefits of recurrent-state transfer and reconstruction conditioning, and corpus-level diagnostics show that the internal branch primarily captures residual rather than target-speech content. These results establish GTCRN-SS as an effective quality--efficiency operating point for causal monaural speech separation.\end{abstract}


\section{Introduction}

Speech separation aims to recover individual speakers from a mixture in which sources overlap in time and frequency, a machine formulation of the cocktail party problem~\citep{Cherry1953}. It is central to hands-free and other single-microphone applications in which concurrent voices must be recovered from a single recording~\citep{WangChen2018}. Although the single-channel setting avoids microphone arrays, it is intrinsically challenging because both source identity and spectral content must be inferred from one waveform. In causal, resource-constrained applications, a separator must further balance separation quality, future-context access, latency, and computational cost~\citep{ReSepFormer2024,Chen2025TFACM}.

Dual-path recurrence and attention have substantially improved speech separation through joint temporal and spectral modeling~\citep{Luo2020DPRNN,Wang2023TFGridNet}. Their standard forms, however, are not inherently causal or computationally lightweight. Without future frames, a separator must preserve speaker identity and interference history through causal recurrent states or cached past context~\citep{Li2022SkiM,Chen2025TFACM}. At the same time, phase-sensitive reconstruction must determine how speech, noise, and uncertain mixture energy are distributed across a fixed set of outputs~\citep{Williamson2015CRM}. These requirements make compact causal separation both a memory problem and an output-allocation problem.

This work investigates how a compact single-output enhancement backbone can be repurposed for noisy, fixed-output separation without future feature frames. We propose GTCRN-SS, a two-stage separator built on GTCRN, whose efficient backbone combines ERB compression, spectral feature expansion, and grouped recurrent modeling~\citep{Rong2024GTCRN}. Rather than treating allocation, state transfer, and refinement as independent additions, GTCRN-SS couples them through a single allocation--refinement contract. Its Gated Complex Mask estimator with a residual sink (GCM\-RS) conserves the complex mixture across the returned speech branches and an internal residual branch. The returned-speech complement of the residual branch conditions Stage~2, whose final masks remain anchored to the original mixture. A frame-aligned low-rank adapter additionally transfers causal grouped-GRNN states across recursive executions without introducing another recurrent memory network.

The neural dependency graph excludes future feature frames; the centered short-time Fourier transform contributes a fixed analysis look-ahead that is reported separately from neural causality. On noisy 8-kHz Libri2Mix, GTCRN-SS achieves 10.1~dB SDRi and 9.5~dB SI-SNRi with 0.9~M parameters and 6.1~G MACs for a one-second input. Independently trained ablations support the contributions of cross-execution state transfer and reconstruction conditioning. Corpus-level sink diagnostics, duration-stratified results, and prefix-invariance tests further characterize residual allocation and causal full-utterance behavior.

Our contributions are threefold:
\begin{itemize}
\item We introduce an allocation--refinement contract that assigns the additional exact complex-allocation branch an internal residual role and uses its returned-speech complement as the Stage~2 reconstruction cue.
\item We propose a frame-aligned, low-rank cross-execution bridge that transforms causal GRNN states and injects them into subsequent GRU gates without requiring another recurrent memory network.
\item We develop a compact two-stage separator and evaluate it through independently trained ablations, corpus-level residual diagnostics, duration-stratified analysis, and full-forward dependency audits.
\end{itemize}

\section{Related Work}
\label{sec:related_work}

In neural separation, permutation-invariant training enabled source-order-independent supervision~\citep{Kolbaek2017uPIT}, while TasNet established end-to-end separation in a learned representation~\citep{TaSNet2018}. Subsequent systems improved long-range and time-frequency modeling through dual-path recurrence and complex-spectrogram processing~\citep{Luo2020DPRNN,Wang2023TFGridNet}. Many high-performing architectures, however, rely on bidirectional or utterance-wide context. Causal systems must instead preserve historical evidence explicitly. SkiM propagates segment-level hidden and cell states through memory LSTMs before initializing the next block~\citep{Li2022SkiM}, and TFACM re-encodes temporal-block states with LSTM-H and LSTM-C modules for cross-block reuse~\citep{Chen2025TFACM}. GTCRN-SS shares the goal of transferring recurrent evidence but uses a non-recurrent, frame-aligned adapter whose output is injected into the gates of the next grouped-GRU execution rather than used only as an initial state.

Multi-stage and cascaded processing are established strategies for improving an initial separation estimate. TasTas refines dual-path estimates while using auxiliary speaker-identity supervision~\citep{Shi2020TasTas}. Neri and Braun study cascades of separately trained noise-suppression, separation, and dereverberation modules, together with subtractive separation~\citep{Neri2023RealTime}. GTCRN-SS therefore does not claim multi-stage processing itself as novel. Its distinction lies in the coupling between stages: both stages are trained under one speech-separation objective, Stage~2 is conditioned by the returned-speech sum from Stage~1, and the final complex masks are applied to the original mixture rather than recursively masking a previous estimate.

Mixture consistency constrains separated outputs to explain the observed mixture. TasNet normalizes source masks with a source-wise softmax in its learned representation~\citep{TaSNet2018}; TF-GridNet uses a soft mixture-constraint loss~\citep{Wang2023TFGridNet}; and differentiable projection can enforce exact consistency for real- or complex-valued estimates~\citep{Wisdom2019MixtureConsistency}. GTCRN-SS builds on this established principle rather than proposing a new consistency projection. Its contribution is the allocation semantics and stage coupling: an additional exact complex branch remains internal as a residual sink, while the complementary sum of the returned speech branches becomes the reconstruction cue for Stage~2.

Resource-efficient models seek a better separation quality--cost trade-off, as illustrated by RE-SepFormer~\citep{ReSepFormer2024}. GTCRN pursues an especially compact operating point for single-output speech enhancement through ERB compression, spectral feature expansion, and grouped recurrence~\citep{Rong2024GTCRN}. Because the original GTCRN has neither a multi-speaker output space nor a permutation-invariant separation objective, it is not directly executable as a separation baseline. GTCRN-SS redesigns this enhancement backbone for fixed-output separation and evaluates the resulting model against causal separation systems under explicit parameter, computation, and runtime accounting.

\section{Proposed Speech Separation}
\label{sec:proposed}

\subsection{Problem Formulation}
\label{subsec:problem_formulation}
\begin{figure*}[!t]
\centering
\includegraphics[width=\textwidth]{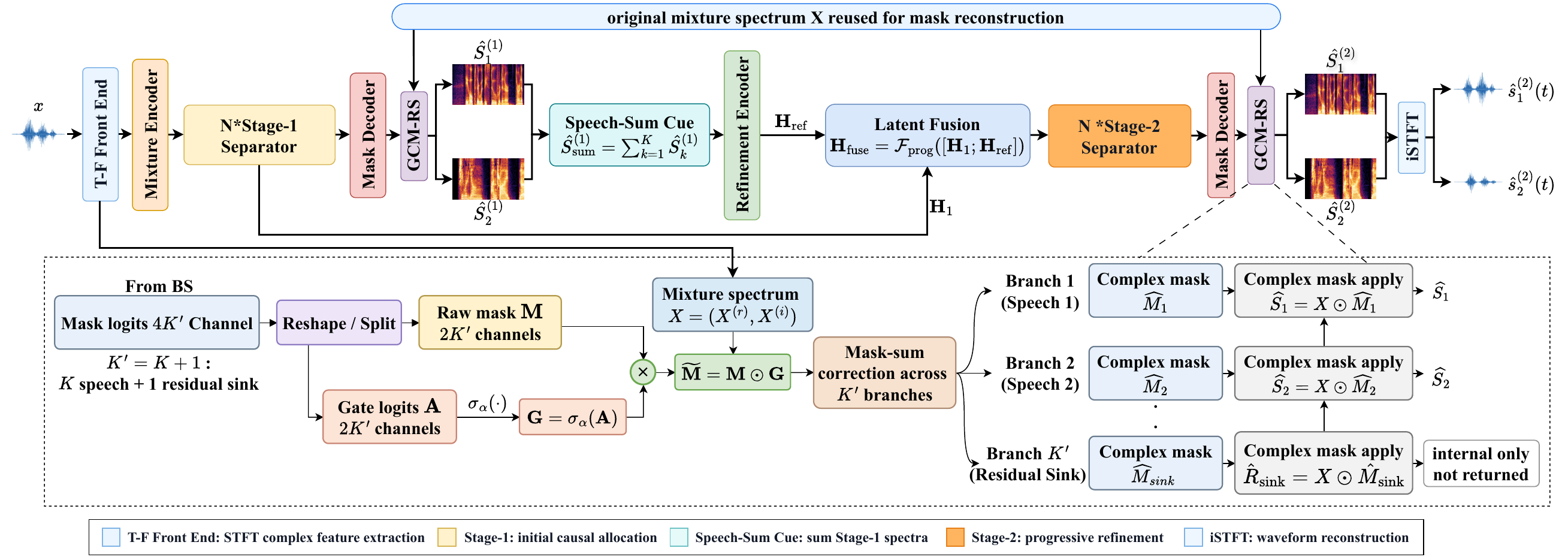}
\caption{Two-stage GTCRN-SS and its residual-allocation GCM-RS head. ERB-SFE maps 129 STFT bins to 65 features, and the strided encoder yields $F_{\mathrm{dp}}=33$. Each head estimates $K$ speech masks plus one internal sink mask; only the speech branches are returned.}
\label{fig:gcm_backbone}
\end{figure*}

Let $x \in \mathbb{R}^{L}$ denote a single-channel mixture,
\begingroup
\setlength{\abovedisplayskip}{2pt plus 1pt minus 1pt}
\setlength{\belowdisplayskip}{2pt plus 1pt minus 1pt}
\setlength{\abovedisplayshortskip}{1pt plus 1pt minus 1pt}
\setlength{\belowdisplayshortskip}{1pt plus 1pt minus 1pt}
\begin{equation}
\label{eq:mixture_definition}
x[n] = \sum_{k=1}^{K} s_k[n] + r[n],
\end{equation}
\endgroup
where $s_k$ is the $k$-th target speech source under the adopted dataset definition and $r$ is any non-returned mixture component (e.g., background noise). The returned speaker set excludes $r$.
Equation~\ref{eq:mixture_definition} fixes the returned-source scope used throughout the model and evaluation.
The network estimates a fixed set of $K$ speech waveforms $\{\hat{s}_1,\ldots,\hat{s}_K\}=f_{\theta}(x)$ and is trained with permutation-invariant assignment.
Within the neural graph, the estimate at feature frame $t$ may depend on current and past feature frames only; Section~\ref{subsec:causal_operation} accounts separately for centered-STFT buffering.
This constraint makes temporal continuity and residual allocation central: speaker identity, voiced segments, and interference remnants must be tracked without future-frame correction.
Figure~\ref{fig:gcm_backbone} shows this two-stage contract.

\subsection{Two-Stage Framework}
\label{subsec:overall_framework}
The model computes a one-sided STFT $X\in\mathbb{C}^{B\times T\times F}$ and stacks magnitude and real-imaginary components into $\mathbf{Z}\in \mathbb{R}^{B \times 3 \times T \times F}$, where $B$, $T$, and $F$ denote batch size, frame count, and one-sided frequency-bin count.
An ERB-domain compression and spectral feature expansion (SFE) front end inherited from GTCRN~\citep{Rong2024GTCRN} compresses $\mathbf{Z}$ to $F_{\mathrm{erb}}$ bands while preserving low-frequency pitch and formant detail; one frequency-downsampling layer then yields the width $F_{\mathrm{dp}}$ processed by the separator.
The mixture encoder applies a frequency-strided convolution and three gated temporal blocks with dilations $d\in\{1,2,5\}$ and causal left-padding $2d$ (no right-padding), producing $\mathbf{H}_{0}\in\mathbb{R}^{B\times C_{\mathrm{hidden}}\times T\times F_{\mathrm{dp}}}$.
Stage~1 passes $\mathbf{H}_{0}$ through the Dual-Path Grouped RNN and Temporal Context Attention (DPGRNN-TCA) separator, decoder, inverse ERB projection, and GCM-RS to obtain preliminary spectra $\{\hat{S}_{k}^{(1)}\}$; Stage~2 processes their speech-sum cue as described in Section~\ref{subsec:stage2_refinement}.
GCM-RS projects each raw gated mask $\tilde M_j$, $j\in\{1,\ldots,K+1\}$, as follows:
\begingroup
\setlength{\abovedisplayskip}{3pt plus 1pt minus 1pt}
\setlength{\belowdisplayskip}{3pt plus 1pt minus 1pt}
\setlength{\abovedisplayshortskip}{2pt plus 1pt minus 1pt}
\setlength{\belowdisplayshortskip}{2pt plus 1pt minus 1pt}
\begin{equation}
\label{eq:mask_projection}
\begin{aligned}
M_{j,\mathrm{R}} &= \tilde M_{j,\mathrm{R}}-
\frac{\sum_l\tilde M_{l,\mathrm{R}}-1}{K+1}, \\
M_{j,\mathrm{I}} &= \tilde M_{j,\mathrm{I}}-
\frac{\sum_l\tilde M_{l,\mathrm{I}}}{K+1}.
\end{aligned}
\end{equation}
\endgroup
Hence $\sum_j M_j=1$ at every time-frequency bin and
$\sum_{k=1}^{K}\hat S_k+\hat R_{\mathrm{sink}}=X$ under complex masking.
The sink is internal and receives no source target; it provides an exact outlet for mixture content not assigned to a returned speaker.
The decoder mirrors the encoder with causal transposed convolutions and past-only temporal padding; each temporal convolution with dilation $d$ uses only frames $t$, $t-d$, and $t-2d$.

\subsection{DPGRNN-TCA Separator}
\label{subsec:separator}
\begin{figure*}[!t]
\centering
\includegraphics[width=1\textwidth]{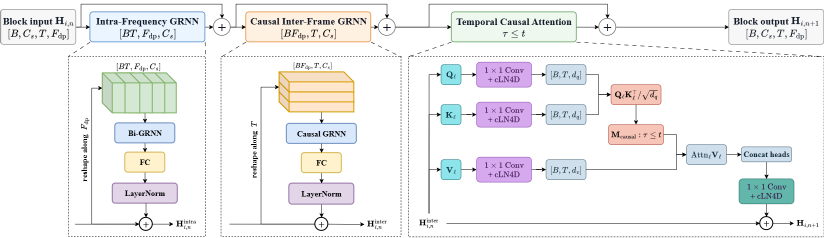}
\caption{One DPGRNN-TCA block at $F_{\mathrm{dp}}=33$. Its 72 channels form two 36-channel groups. Bidirectional recurrence is restricted to within-frame frequency; temporal recurrence is unidirectional, the CE bridge is frame-aligned, and TCA uses a lower-triangular mask.}
\label{fig:dpgrnn_tca_block}
\end{figure*}

Both stages use separate instances of the same recursive design: $N=3$ DPGRNN-TCA blocks, each executed $I=3$ times.
As Figure~\ref{fig:dpgrnn_tca_block} shows, each block applies intra-frequency GRNN, causal inter-frame GRNN, and temporal context attention.
Normalization never aggregates future frames: recurrent branches use per-frame LayerNorm over frequency and channel, whereas convolutional and attention paths use cumulative statistics through frame $t$.

Within each block, the input is first reshaped to $(BT)\times F_{\mathrm{dp}}\times C_{\mathrm{hidden}}$. A bidirectional grouped GRNN~\citep{Gao2018GroupRNN} processes each frame's frequency-bin sequence, followed by linear projection, layer normalization~\citep{Ba2016LayerNorm}, and residual addition.
Bidirectionality is confined to frequency and accesses no future frames.

The tensor is then reshaped to $(BF_{\mathrm{dp}})\times T\times C_{\mathrm{hidden}}$, giving each frequency bin its own temporal sequence. A grouped causal GRNN propagates speaker and interference history along this sequence.

\begin{figure}[!t]
\centering
\includegraphics[width=\columnwidth]{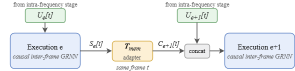}
\vspace{-15pt}
\caption{Cross-execution (CE) bridge. A pointwise low-rank residual adapter transforms $\mathbf{S}_{e}[t]$ into $\mathbf{C}_{e+1}[t]$, which conditions the next execution's GRU gates at frame $t$ rather than initializing its hidden state.}
\label{fig:ce_bridge}
\end{figure}

Across recursive executions, the cross-execution state bridge (CE bridge) reuses recurrent evidence from different latent inputs.
Figure~\ref{fig:ce_bridge} illustrates the frame-aligned state-transfer path.
Let $e$ index successive block executions.
The causal inter-frame GRNN produces a state sequence $\mathbf{S}_{e}$, where $\mathbf{S}_{e}[t]$ summarizes frames up to $t$.
A stage-specific low-rank residual adapter then computes
\begin{equation}
\mathbf{C}_{e+1}
=
\mathrm{LN}\!\left(
\mathbf{S}_{e}
+
\mathbf{W}_{\uparrow}
\mathrm{SiLU}\!\left(
\mathbf{W}_{\downarrow}\mathbf{S}_{e}
\right)
\right),
\label{eq:hidden_transition}
\end{equation}
where $\mathbf{W}_{\downarrow}\in\mathbb{R}^{(C_{\mathrm{hidden}}/4)\times C_{\mathrm{hidden}}}$ and $\mathbf{W}_{\uparrow}\in\mathbb{R}^{C_{\mathrm{hidden}}\times(C_{\mathrm{hidden}}/4)}$ are pointwise channel projections.
The adapted sequence $\mathbf{C}_{e+1}$ is concatenated into the grouped GRU gate input of the next execution.
The bridge is frame-aligned: $\mathbf{C}_{e+1}[t]$ contains only information already summarized in $\mathbf{S}_{e}[t]$ and conditions the next execution at the same frame.
It is not used as that execution's initial hidden state.
The adapter itself has no temporal recurrence; temporal history remains in the causal inter-frame GRNN state.
No bridge is applied after the final execution of a stage.
One bridge module is shared within each stage; the two stages use separate parameters.

Finally, temporal context attention (TCA) uses causal scaled dot-product attention~\citep{Vaswani2017Attention}. Query, key, and value features are produced by $1\times1$ convolutions and flattened across channel and frequency. A lower-triangular additive mask prevents frame $t$ from attending to any $\tau>t$. With two heads and $C_q=4$, each head has query dimension $C_qF_{\mathrm{dp}}=132$. TCA refines the observed prefix, while primary temporal history is carried by the inter-frame GRNN states and CE bridge.

\subsection{Causal Operation}
\label{subsec:causal_operation}
Following the real-time causal formulation of TFACM~\citep{Chen2025TFACM}, computation at frame $t$ uses current and preceding feature frames but no future frame. This follows from unidirectional recurrence, masked TCA, causal normalization, and past-only padding. The centered STFT ($W=256$, $H=64$ at 8~kHz) still adds 16~ms of analysis look-ahead. Because TCA retains the observed prefix, its cost grows with utterance length. We therefore report one-second MACs, 1--8-s batch-one inference times, and a prefix audit.

\subsection{Reconstruction-Conditioned Processing}
\label{subsec:stage2_refinement}
A compact causal pass may leave residual interference that cannot be revised with future context.
Stage~2 therefore revisits the Stage-1 estimate without redefining the output space.
The Stage-1 speech spectra are summed as $\hat{S}_{\mathrm{sum}}^{(1)}=\sum_{k=1}^{K}\hat{S}_{k}^{(1)}$, forming a speaker-combined returned-branch cue that collapses the unordered speech branches without assigning it a speaker index.
Magnitude and RI features recomputed from $\hat{S}_{\mathrm{sum}}^{(1)}$ pass through the same ERB-SFE front end and a refinement encoder (same architecture as the mixture encoder, independent parameters) to give $\mathbf{H}_{\mathrm{ref}}$. The Stage-1 latent $\mathbf{H}_{1}$ and reconstruction latent $\mathbf{H}_{\mathrm{ref}}$ are concatenated and mapped to $\mathbf{H}_{\mathrm{fuse}}$ by a $1\times1$ convolution followed by PReLU.
Stage~2 processes the fused latent with an independent separator and decoder.
Its GCM-RS again applies the resulting complex masks to the original mixture spectrum $X$, so the final waveform remains anchored to that mixture.

\section{Experimental Setup}
\label{sec:experiment}

\subsection{Dataset and Preprocessing}
\label{subsec:datasets}
We evaluate on the 8-kHz Libri2Mix~\citep{Cosentino2020LibriMix} \textit{train-100/min/mix\_both} condition, which contains two-speaker mixtures with WHAM! noise~\citep{Wichern2019WHAM}. Training and validation use the first 3~s of each released fixed mixture, following the public TFACM audio-only recipe~\citep{Chen2025TFACM}. We apply no random cropping, dynamic mixing, or additional data augmentation. Evaluation uses all 3,000 test utterances at their complete released durations. Of these, 2,987 exceed 3~s; the test set therefore evaluates predominantly longer inputs than those used during training.

\subsection{Model Configuration}
\label{subsec:training}
The STFT uses a 256-sample Hann window and a 64-sample hop, corresponding to a 32-ms window and an 8-ms frame update. The 129 one-sided frequency bins are mapped to 65 ERB-SFE features and then downsampled to $F_{\mathrm{dp}}=33$. Each stage has a hidden width of 72 channels, three DPGRNN-TCA blocks, three recursive executions, two GRNN groups, and two attention heads. The Stage-2 fusion layer maps the concatenated 144-channel representation back to 72 channels.

\subsection{Training Objective and Optimization}
Following TFACM, we train with utterance-level permutation-invariant training (PIT)~\citep{Kolbaek2017uPIT,Chen2025TFACM}. For a reference waveform $s$ and estimate $\hat{s}$, let $\bar{z}=z-\mathrm{mean}(z)$. The pairwise scale-sensitive waveform score is
\begin{equation}
\label{eq:snr_objective}
\mathrm{SNR}_{\epsilon}
=
10\,\log_{10}
\left(
\frac{\|\bar{s}\|^{2}}
{\|\bar{\hat{s}}-\bar{s}\|^{2}+\epsilon}
+\epsilon
\right),
\end{equation}
with $\epsilon=10^{-8}$.
PIT selects the source assignment that maximizes the utterance-level score over the complete 3-s training segment. For each supervised stage, an auxiliary term $\mathcal{L}_{\mathrm{fw}}$ adds equally weighted waveform and real/imaginary spectral L1 errors, giving $\mathcal{L}^{(q)}=-\mathrm{SNR}_{\epsilon}^{(q)}+\mathcal{L}_{\mathrm{fw}}^{(q)}$ for $q\in\{1,2\}$. Both stage outputs use the same signal-level terms. The total training objective is
\begin{equation}
\label{eq:training_objective}
\mathcal{L}_{\mathrm{train}} = 0.3\,\mathcal{L}^{(1)} + 1.0\,\mathcal{L}^{(2)},
\end{equation}
which supervises both stages while emphasizing the final output. The internal sink receives no independent residual target.

All models are implemented in PyTorch~2.6.0+cu118 and optimized with Adam for at most 400 epochs. We use an initial learning rate of $10^{-3}$, weight decay of $10^{-4}$, and gradient-norm clipping at 5.0. The full model and the three comparison systems in Table~\ref{tab:mechanism} are independently initialized and trained on the same split and fixed 3-s segments with the same objective and optimization settings. Variant B sets the CE signal to zero throughout training and evaluation; variant C similarly zeros the Stage-1 reconstruction cue; variant D replaces the residual-allocation head with exact $K$-branch consistency throughout training and evaluation.

\begin{table}[!t]
\centering
\footnotesize
\setlength{\tabcolsep}{3.2pt}
\renewcommand{\arraystretch}{1.08}
\begin{tabular}{@{}lrrrr@{}}
\toprule
Method
& \multicolumn{1}{c}{\shortstack{SDRi\\(dB)}}
& \multicolumn{1}{c}{\shortstack{SI-SNRi\\(dB)}}
& \multicolumn{1}{c}{\shortstack{Params\\(M)}}
& \multicolumn{1}{c}{\shortstack{MACs\\(G)}} \\
\midrule
DPRNN & 9.6 & 9.0 & 1.3 & 11.2 \\
SkiM & 9.9 & 9.2 & 3.6 & 11.1 \\
ReSepFormer & 4.1 & 3.3 & 8.0 & 6.3 \\
TF-GridNet-Causal & 11.5 & 10.8 & 11.4 & 179.2 \\
TFACM (Small) & 10.2 & 9.3 & 0.5 & 19.4 \\
\midrule
\textbf{GTCRN-SS (ours)} & 10.1 & 9.5 & 0.9 & 6.1 \\
\bottomrule
\end{tabular}
\vspace{-5pt}
\caption{Quality--cost comparison for causal 8-kHz separation. Higher SDRi and SI-SNRi are better; lower parameter and MAC counts are better. Reference rows follow~\citet{Chen2025TFACM}.}
\label{tab:sota}
\end{table}

\subsection{Evaluation Protocol and Hardware}
\label{subsec:evaluation_protocol}

For separation quality, we use SI-SNRi as the primary metric and SDRi as a corroborative metric~\citep{Vincent2006BSS,LeRoux2019SDR}. Both follow the two-speaker PIT convention implemented by Look2Hear and adopted by TFACM~\citep{Chen2025TFACM}. Reported scores are unweighted means over the 3,000 complete test utterances. Paired intervals are obtained from 50,000 utterance-level bootstrap resamples of the aligned outputs from each model pair.

For model complexity, we report total trainable parameters and multiply--accumulate operations (MACs). GTCRN-SS contains 0.9~M trainable parameters. Following the public TFACM complexity protocol~\citep{Chen2025TFACM}, MACs are measured with ptflops~0.7.5 in evaluation mode for a fixed one-second 8-kHz input. Values for external systems in Table~\ref{tab:sota} are retained from their cited reports.

For runtime, we benchmark local GTCRN-SS and TFACM-Small implementations using batch-one 8-kHz waveforms of 1, 2, 4, and 8~s. Measurements use PyTorch~2.6.0+cu118 on a workstation with an NVIDIA GeForce RTX~3060 (12~GB) and an AMD Ryzen~5 7500F; CPU runs are restricted to one thread. GPU results contain 100 timed trials after 10 warm-up trials, and CPU results contain 20 timed trials after 5 warm-up trials. We report the median and empirical 5th--95th percentiles at each device--duration setting.

\section{Results and Analysis}
\label{sec:results}

We first place GTCRN-SS in the reported quality--cost landscape and then compare runtime under a shared implementation environment. Controlled ablations evaluate the proposed information paths, while residual-sink, stage-progression, duration, and prefix analyses characterize model behavior beyond aggregate separation scores.

\subsection{Comparison with Causal Systems}
\label{subsec:sota}

Table~\ref{tab:sota} compares the measured GTCRN-SS operating point with reported causal 8-kHz separation systems~\citep{Luo2020DPRNN,Li2022SkiM,ReSepFormer2024,Chen2025TFACM}. Reference rows follow the TFACM report, whereas the GTCRN-SS row is measured under our Libri2Mix condition and implementation. Because these rows do not share a controlled training and evaluation pipeline, the table is used to establish external quality--cost context rather than pairwise causal attribution.

GTCRN-SS reaches 10.1~dB SDRi and 9.5~dB SI-SNRi with 0.9~M parameters and 6.1~G MACs. It has the lowest MAC count among the listed systems and improves SI-SNRi over the other low-compute entries. TF-GridNet-Causal retains the highest separation scores but requires substantially more parameters and computation. Controlled evidence for the proposed components is provided by the independently trained comparisons below.

\begin{figure}[!t]
\centering
\includegraphics[width=0.9\columnwidth]{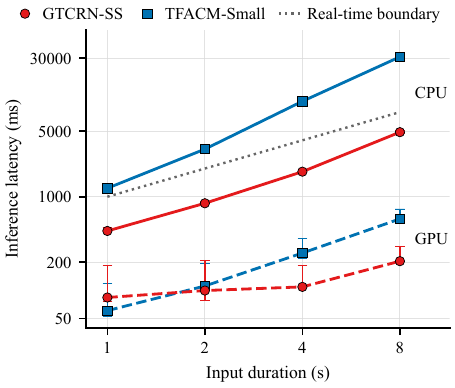}
\vspace{-5pt}
\caption{Same-host batch-one inference-time scaling. Solid and dashed lines denote CPU and GPU runs, respectively; markers and bars show medians and empirical 5th--95th percentiles.}
\label{fig:runtime_scaling}
\vspace{-5pt}
\end{figure}

\subsection{Causal Efficiency and Same-Host Runtime}
\label{subsec:runtime_profile}

Figure~\ref{fig:runtime_scaling} compares GTCRN-SS and TFACM-Small using the same-host protocol in Section~\ref{subsec:evaluation_protocol}. This comparison complements the one-second MAC counts with measured full-utterance latency.

For a 1-s input on the GPU, GTCRN-SS is slower than TFACM-Small (84.0 vs.\ 60.9~ms). It is faster at the remaining seven device--duration settings. At 8~s, the GTCRN-SS median is 205.2~ms on the GPU and 4{,}888.4~ms on the CPU, compared with 579.9 and 31{,}121.6~ms for TFACM-Small. These correspond to 2.8$\times$ and 6.4$\times$ lower latency, respectively. GTCRN-SS maintains a CPU real-time factor below one through 8~s, whereas TFACM-Small exceeds one at every measured duration. These measurements characterize full-utterance throughput, not per-hop streaming latency.

\begin{table*}[!t]
\centering
\footnotesize
\setlength{\tabcolsep}{5pt}
\renewcommand{\arraystretch}{1.10}
\begin{tabular*}{\textwidth}{@{\extracolsep{\fill}}clcccrr@{}}   
\toprule
& & \multicolumn{3}{c}{Information and allocation} & \multicolumn{2}{c}{Performance (dB)} \\
\cmidrule(lr){3-5}\cmidrule(l){6-7}
ID & Configuration & CE input & Stage-2 cue & Sink
& \multicolumn{1}{c}{SI-SNRi} & \multicolumn{1}{c}{SDRi} \\
\midrule
A & \textbf{Full model} & state & speech sum & yes & \textbf{9.5} & \textbf{10.1} \\
B & Without CE information & zero & speech sum & yes & 8.1 & 8.8 \\
C & Without reconstruction cue & state & zero & yes & 7.7 & 8.4 \\
\midrule
D$^\dagger$ & $K$-branch exact consistency & state & mixture $X$ & no & 6.1 & 6.4 \\
\bottomrule
\end{tabular*}                                                     
\vspace{-5pt}
\caption{Independently trained ablations. B removes CE information, C removes the reconstruction cue, and D$^\dagger$ replaces the complete $(K+1)$-branch allocation--conditioning contract with exact $K$-branch consistency. Each modification is active during both training and evaluation.}
\label{tab:mechanism}
\end{table*}

\subsection{Information-Path Ablations}
\label{subsec:ablation}

Table~\ref{tab:mechanism} reports independently trained comparisons. Variant B removes CE information, variant C removes the Stage-1 reconstruction cue, and variant D$^\dagger$ replaces the complete residual-allocation contract with exact $K$-branch consistency. Each modification is active throughout training and evaluation. Removing CE information reduces SI-SNRi by 1.40~dB (95\% paired interval: [1.27, 1.52]) and SDRi by 1.32~dB ([1.19, 1.44]). Removing the reconstruction cue produces larger reductions of 1.75~dB SI-SNRi ([1.62, 1.89]) and 1.67~dB SDRi ([1.54, 1.80]). These independently trained comparisons show that both cross-execution state transfer and reconstruction conditioning contribute under the shared training protocol; they do not establish that the particular bridge or cue function is optimal among alternatives.

Variant D$^\dagger$ reaches 6.1~dB SI-SNRi and 6.4~dB SDRi. Relative to the full model, the decreases are 3.36~dB ([3.24, 3.48]) and 3.65~dB ([3.54, 3.77]), respectively. This comparison changes both output allocation and Stage-2 conditioning: exact $K$-branch consistency removes the internal residual outlet and makes the returned Stage-1 sum equal to the mixture. The result therefore evaluates the complete allocation--conditioning contract rather than an isolated sink-only effect.

\subsection{Residual-Allocation Behavior}
\label{subsec:residual_behavior}

\begin{figure*}[!t]
\centering
\includegraphics[width=0.9\textwidth]{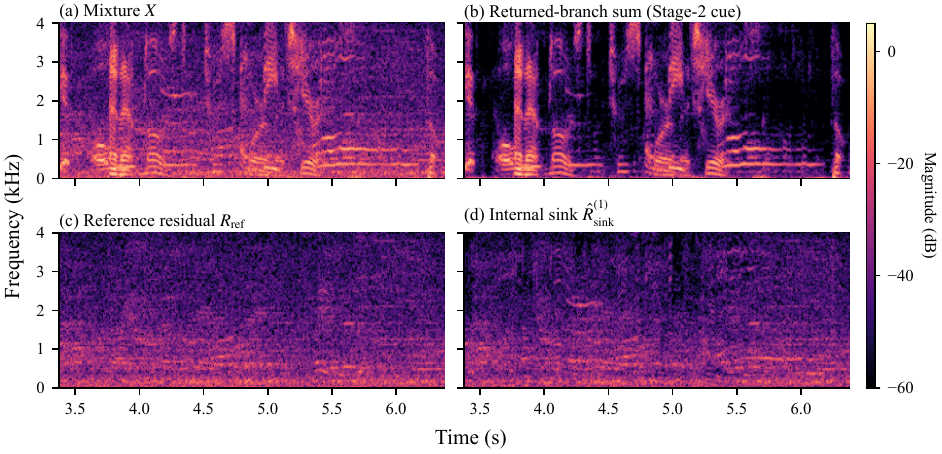}
\vspace{-10pt}
\caption{Stage-1 residual allocation: (a) mixture, (b) returned sum used by Stage~2, (c) reference residual, and (d) internal sink. All panels share one mixture-referenced dB scale.}
\label{fig:residual_sink_allocation}
\end{figure*}

Figure~\ref{fig:residual_sink_allocation} visualizes the test utterance closest to the full model's aggregate SI-SNRi and SDRi. All panels use the common mixture-referenced scale $20\log_{10}(|Y|/\max|X|)$ clipped to $[-60,5]$~dB. Before visualization, an inference hook verifies the Stage-1 allocation identity $\hat S_{\mathrm{sum}}^{(1)}+\hat R_{\mathrm{sink}}^{(1)}=X$.

The identity follows from the projection in Equation~\ref{eq:mask_projection}; the figure illustrates how the unsupervised internal branch distributes energy relative to the released residual reference. The following corpus-level analysis tests whether this behavior generalizes beyond one example.

\subsection{Corpus Residual-Sink Alignment}
\label{subsec:sink_corpus}

For every complete test mixture, we reconstruct the internal Stage-1 sink and define the released waveform residual as $r_{\mathrm{ref}}=x-s_1-s_2$. The implemented allocation identity has a maximum absolute numerical error of $3.5\times10^{-7}$. Table~\ref{tab:sink_semantics} compares absolute Pearson correlations between the centered sink waveform and the residual or speech references. The reported intervals use 50,000 utterance-level bootstrap resamples while holding the trained full-model checkpoint fixed.

\begin{table}[!t]
\centering
\footnotesize
\setlength{\tabcolsep}{5pt}
\renewcommand{\arraystretch}{1.10}
\begin{tabular}{@{}lcc@{}}
\toprule
Metric & Estimate & 95\% interval \\
\midrule
Residual correlation & 0.843 & [0.840, 0.845] \\
Max. speech correlation & 0.139 & [0.137, 0.140] \\
Correlation advantage & 0.704 & [0.700, 0.708] \\
Residual-preferred utterances & 99.9\% & --- \\
\bottomrule
\end{tabular}
\vspace{-5pt}
\caption{Residual-sink alignment over 3,000 complete test mixtures. Correlation estimates are mean absolute centered-waveform Pearson correlations; intervals are 95\% utterance-level bootstrap intervals.}
\label{tab:sink_semantics}
\end{table}

The sink is more strongly correlated with the released residual than with either speech source for 99.9\% of utterances. Its mean absolute residual correlation is 0.843, compared with 0.139 for the larger speech correlation, yielding an average advantage of 0.704. This result supports residual-aligned allocation by the trained model. Because the sink receives no residual supervision, however, it should not be interpreted as a directly supervised noise estimate.

\subsection{Stage Progression}
\label{subsec:stage_prefix_analysis}

Within the full model, Stage~1 obtains 8.5~dB SI-SNRi and 9.2~dB SDRi, whereas Stage~2 reaches 9.5 and 10.1~dB. The paired improvements are 0.96~dB SI-SNRi (95\% interval: [0.93, 0.99]) and 0.92~dB SDRi ([0.89, 0.95]). Stage~2 improves 88.9\% of test utterances in SI-SNRi and 88.3\% in SDRi. These within-model results show that the second stage usually improves the supervised intermediate estimate; together with variant C, they also show that reconstruction information is useful to the trained two-stage architecture.

We also examine full-utterance duration behavior. Of the 3,000 test mixtures, 2,987 exceed the 3-s training duration. Across four equal-count duration ranges spanning 3.00--13.99~s, SI-SNRi remains between 9.3 and 9.6~dB and SDRi remains between 9.9 and 10.3~dB. We observe no monotonic degradation with duration under this test condition.

For the prefix-invariance audit, waveform pairs share an identical prefix and have independently sampled suffixes; splice points are aligned to the 8-ms hop. We compare both model stages before guarded splice boundaries across five suffix seeds, two durations, and three prefix fractions, producing 60 stage-level comparisons. All comparisons satisfy the declared $10^{-5}$ absolute and $10^{-6}$ relative-RMS tolerances for guards of at least 24~ms, the smallest passing guard tested. The centered STFT contributes 16~ms of feature-analysis look-ahead, while the additional 8-ms guard excludes overlap-add samples adjacent to the artificial splice. Thus, 16~ms describes transform buffering and 24~ms describes the minimum passing boundary in this finite audit; neither is a measurement of per-hop streaming latency.

\section{Discussion}

The results support GTCRN-SS as a coupled allocation--refinement design. The internal branch enforces mixture conservation and provides a permutation-invariant complement for Stage~2. Its strong residual correlation supports the intended allocation behavior, but does not make the unsupervised sink a supervised noise estimate. The independently trained ablations further support cross-execution state transfer and reconstruction conditioning, without establishing that the proposed bridge or cue is optimal. Because variant D changes both allocation and Stage-2 conditioning, it should be interpreted as a contract-level comparison rather than a sink-only ablation.

The model provides a favorable quality--efficiency operating point for causal full-utterance inference. However, real-time factor and MAC count do not establish streaming latency or bounded memory: the centered STFT introduces fixed look-ahead, and full-prefix TCA has length-dependent cost. Claims are therefore limited to the evaluated full-utterance setting.

The evaluation is restricted to noisy two-speaker 8-kHz Libri2Mix. Bootstrap intervals condition on fixed checkpoints and do not capture variation across training runs, while cross-paper comparisons differ in experimental conditions. Broader acoustic settings, repeated training seeds, and bounded-context streaming evaluation remain necessary.

\section{Conclusion and Future Work}
\label{sec:conclusion}

We introduced GTCRN-SS, a compact two-stage separator that combines internal residual allocation, reconstruction-conditioned refinement, and frame-aligned recurrent-state transfer. On noisy 8-kHz Libri2Mix, it achieves 10.1~dB SDRi and 9.5~dB SI-SNRi with 0.9~M parameters and 6.1~G MACs. Independently trained ablations support both information paths, while residual and prefix diagnostics characterize the allocation behavior and causal scope of the evaluated full-utterance system. Future work will evaluate variability across training seeds and broader speaker and acoustic conditions. We will investigate bounded-context or cached attention for constant-memory streaming and measure per-hop latency on resource-constrained devices.

\clearpage
\bibliography{rocling2026}
\bibliographystyle{acl_natbib}

\end{document}